\documentclass[fleqn,twoside]{article}
\usepackage{espcrc2}

\usepackage{graphicx}
\usepackage[figuresright]{rotating}

\newcommand{\AmS}{{\protect\the\textfont2
  A\kern-.1667em\lower.5ex\hbox{M}\kern-.125emS}}

\title{Beauty and charm to study new physics at future linear colliders}

\author{M. Battaglia\address{Santa Cruz Institute of Particle Physics, 
University of California at Santa Cruz, CA 95064, USA},
\address{Lawrence Berkeley National Laboratory, Berkeley, CA 94720, USA},
\address{CERN, DG Department, CH-1211 Geneva, Switzerland}}%
\begin{document}

\begin{abstract}
$b$ and $c$ hadrons are instrumental to the identification and study of the 
Higgs sector and new physics at a future lepton collider. This paper reviews 
highlights of $b$ and $c$ physics for the linear collider programs and the 
directions of ongoing R\&D on pixellated Si sensors for its vertex tracker.

\vspace{1pc}
\end{abstract}

\maketitle

\section{INTRODUCTION}

An $e^+e^-$ linear collider (LC) has emerged as possibly the most practical and 
realistic way towards collisions of elementary particles at constituent energies matching 
those of the LHC with high luminosity. The attainable beam energy depends on the 
available accelerating gradient and the luminosity on the beam power and its emittance 
at the interaction point (IP). Two projects, which aim at different collision energies
with different acceleration technologies, are presently being developed. The ILC project 
is based on the use of superconducting RF cavities providing gradients of $\simeq$~30~MV/m 
to produce collisions at centre-of-mass energies $\sqrt{s}$ = 0.25 - 
$\simeq$~1~TeV~\cite{Brau:2007zza}. In order to achieve higher energies the CLIC project 
develops a new acceleration scheme where a low-energy, high current drive beam is used to 
accelerate the main beam through high-frequency transfer structures, which have achieved 
gradients of $\simeq$~100 MV/m~\cite{Assmann:2000hg}.
In a farther future, plasma wake-fields accelerators using high-power 
laser pulses can provide gradients of 1-10 GV/m~\cite{Geddes:2004tb,Thomas:2007jg}.

Heavy flavour identification and decay reconstruction is essential for the 
physics program of the next generation of high energy $e^+e^-$ colliders.
Because $b$ and $t$ are the two heaviest quarks, they will be preferentially 
produced by  particles with large couplings to massive fermions including, 
but not limited to, the Higgs boson(s). Because $b$, $c$ and $t$ can, in principle, 
be identified with high efficiency and purity, they will also enable the selection 
of well-defined exclusive hadronic final states, for example in the study of precision 
electro-weak observables. This paper discusses the role of heavy hadron identification 
and reconstruction in the framework of the linear collider physics program and 
presents some highlights of the R\&D program towards the next generation of vertex 
trackers, matched to the linear collider physics requirements.

\section{HEAVY FLAVOURS AND THE LINEAR COLLIDER PHYSICS PROGRAM}

\begin{figure}[htb]
\includegraphics[width=14pc]{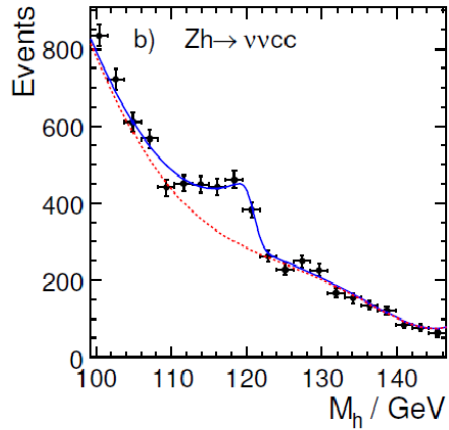}
\vspace*{-1.0cm}
\caption{$c \bar c$ invariant mass distribution showing the $e^+e^- \to Z^0 h^0 \to 
\nu \bar \nu c \bar c$ signal above the SM background for $M_h$ = 120~GeV at $\sqrt{s}$ = 
250~GeV. A $c$-tagging efficiency of 29\% is obtained. The result is obtained with full 
simulation and reconstruction of the ILD detector concept at ILC (from~\cite{ildloi}).}
\label{fig:mhcc}
\end{figure}

The detailed study of the Higgs profile is possibly the centre-piece of the
$e^+e^-$ linear collider physics program at $\sqrt{s}$ = 0.25-0.5~TeV, in 
particular if the Higgs boson, $h^0$, is light, i.e.\ its mass is below the 
$W^+W^-$ threshold, so that its dominant decay is 
$h^0 \to b \bar b$~\cite{Battaglia:2000jb}. Not only tagging the $b$ jets is 
instrumental for studying its production, even more importantly the identification 
of $b$, $c$ and light hadrons in its decay products is an essential test of
the Higgs mechanism of mass generation. If this is indeed responsible for
generating the fermion masses, in addition to those of the gauge bosons, the 
couplings of the Higgs boson to $b$ and $c$ quarks must scale proportionally to
their masses. If the Higgs sector is embedded into an extended model of New 
Physics, such as Supersymmetry (SUSY), these relations receive important 
corrections. Couplings to up-like and down-like fermions are shifted compared to
their Standard Model (SM) predictions. 
In addition, in SUSY models sbottom-gluino and stop-higgsino 
loops may shift the effective $b$-quark mass in the $hbb$ coupling. It is therefore 
essential to accurately determine these couplings by measuring the branching fractions 
of the Higgs boson decays in the corresponding fermions pairs. From $M_H$ = 115~GeV, 
just above the LEP-2 lower mass limit, to $M_H$ = 185~GeV, the SM upper mass limit from 
precision electro-weak observables, the branching fraction BR($H^0 \to b \bar b$) varies 
from 0.71 to 3.8$\times$10$^{-3}$, i.e.\ from being the dominant process to becoming 
a rare decay. For a light Higgs boson, with mass $\simeq$~120~GeV the accessible fermionic 
final states are $b \bar b$, $c \bar c$ and $\tau^+ \tau^-$. These decays need to be 
distinguished among them and from the top-loop mediated $H \to gg$ decay, which yields 
jets with light hadrons. This motivates by efficient and pure jet flavour tagging, which
is best done by using a topological reconstruction of the decay chain~\cite{Bailey:2009ui}. 
Detailed studies have demonstrated that the effective couplings 
of an 120~GeV Higgs boson to $b \bar b$, $c \bar c$ and $gg$ can be measured with a 
relative statistical accuracy of 0.005, 0.06 and 0.04, respectively, with 250~fb$^{-1}$ 
of statistics collected at $\sqrt{s}$ = 250~GeV~\cite{Kuhl:2007zza}. 
Figure~\ref{fig:mhcc} shows the Higgs mass peak obtained in the $c \bar c$ channel with the 
ILD detector concept~\cite{ildloi}.
As the Higgs mass increases, the $W^+W^-$ process becomes dominant. 
Still, measuring the $b \bar b$ final state is essential for verifying the fermion mass 
generation mechanism.
Given the low event rate, this measurement is best performed at $\sqrt{s}$ = 1~TeV, 
or above, where the $e^+e^- \to \nu \bar \nu H^0$ fusion production process offers 
larger cross section compared to the associated production with the $Z^0$, typically 
exploited at $\sqrt{s}$ = 250 - 500~GeV~\cite{Battaglia:2002gq,Barklow:2003hz}. The 
main challenge presented by this measurement is the identification of relatively soft 
$b$ jets in the forward region, i.e.\  at polar angles below 25$^{\circ}$. 

\begin{figure}[htb]
\includegraphics[width=14pc,angle=-90]{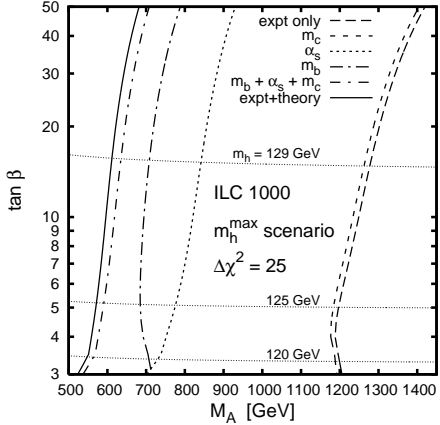}
\vspace*{-0.75cm}
\caption{Contributions of different sources of parametric uncertainties to the 
sensitivity to $M_A$ from measurements of the $h^0$ couplings in SUSY 
(from~\cite{Droll:2006se}).}
\label{fig:hb}
\end{figure}

\begin{figure}[htb]
\includegraphics[width=17pc]{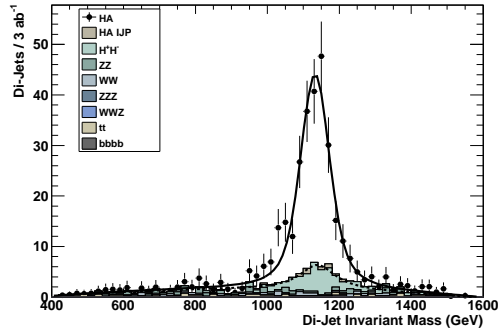}
\vspace*{-0.75cm}
\caption{$b \bar b$ invariant mass distribution showing the $e^+e^- \to H^0 A^0 \to 
b \bar b b \bar b$ signal above the SM and SUSY backgrounds for $M_A$ = 1.14~TeV at 
$\sqrt{s}$ = 3~TeV and assuming a $b$-tagging efficiency $\epsilon_b$=0.85. The result is 
obtained with full simulation and reconstruction of a detector at CLIC 
(from~\cite{Battaglia:2010in}).}
\label{fig:ma}
\end{figure}

Measurements of Higgs couplings to fermions accurate enough to be sensitive to the 
mass scale of heavy states, such as the SUSY Higgs sector, require not 
only high statistical accuracy but also precise theory predictions. Then, the 
interpretation of these measurements crucially depends on the precision of inputs, 
such as $m_b$, $m_c$ and $\alpha_s$, currently being obtained by lower energy 
accelerator experiments~\cite{Droll:2006se}. Figure~\ref{fig:hb} shows the change 
in sensitivity to the mass of the supersymmetric CP-odd $A^0$ boson when assuming 
only the experimental accuracies and also adding the uncertainties on heavy quark 
masses and $\alpha_s$.
If the linear collider $\sqrt{s}$ energy is sufficient, heavy Higgs bosons from SUSY, 
or other models with an extended Higgs sector, can be directly pair produced. 
These bosons are expected to couple predominantly to $t$ and $b$ quarks and their study 
would represent a genuine feast for $b$-tagging and reconstruction. 
The CP-odd and the heavy CP-even neutral Higgs bosons, $A^0$ and $H^0$ give mostly 
$e^+e^- \to H^0 A^0 \to b \bar b b \bar b$ and $b \bar b \tau^+ \tau^- $, while charged 
bosons would predominantly give $e^+e^- \to H^+ H^- \to t \bar b \bar t b$.
The study of these processes is of special relevance for establishing the connection between 
particle physics and cosmology through dark matter. In fact, the $A^0$ boson plays a special 
role in the study of the relic density of Supersymmetric dark matter in the universe. 
For a precise study of dark matter, the observation of the heavy Higgs bosons at colliders 
and the measurement of their properties is essential. 
If $M_A \simeq 2 M_{\chi^0_1}$, the neutralino annihilation process in the early universe is 
significantly enhanced through the $A^0$ pole, $\chi^0_1 \chi^0_1 \to A^0 \to b \bar b$, which 
results in a large reduction of its relic density, $\Omega_{\chi}$. For the relic density 
calculation, the measurement of the $A^0$ mass and width removes a major source of uncertainty. 
For dark matter direct detection, these are equally important for calculating the scattering 
cross section, since the dominant contribution comes most often from Higgs boson exchange diagrams.
The analysis of $H^0A^0$ has to identify the 4-$b$ jets with high efficiency, since the 
expected cross section is $\cal{O} \mathrm{(1~fb)}$ and, with four jets to tag, 
the efficiency is $\propto \epsilon_b^4$, and small mis-identification, since the expected 
background-to-signal ratio is $\cal{O}\mathrm{(5 \times 10^3)}$. A linear collider is 
expected to directly observe these states almost up to the kinematical limit. Detailed 
studies have been performed assuming various collider energies and boson masses. For 
boson masses in the range 400 $< M_A <$ 1100~GeV, an $\epsilon_b$ efficiency of $\simeq$~0.80,
or more, is desirable (see Figure~\ref{fig:ma}).

SUSY loop contributions from $\tilde t$, $\tilde b$ and $\tilde g$ may induce sizable CP 
asymmetries in charged Higgs boson decays~\cite{Christova:2002ke}. These can be observed by 
studying the observable $\delta CP = \frac{\Gamma(H^- \to b \bar t) - \Gamma(H^+ \to \bar b t)}
{\Gamma(H^- \to b \bar t) + \Gamma(H^+ \to \bar b t)}$. For this measurement the heavy quark 
must be distinguished from its anti-quark. This can be done either by looking 
at the charge of the leptons from semileptonic decays or from the vertex charge, which requires
excellent control of the reconstruction of charged decay products~\cite{Bailey:2009ui}. 

Finally, a linear collider of sufficient energy offers a unique opportunity to study the 
Higgs potential through the measurement of the Higgs self-coupling. Simulation studies have shown 
that the role of an hadron collider in this study is marginal, unless it reaches several 
ab$^{-1}$ of integrated luminosity or collision energies of 
$\cal{O} \mathrm{(100~GeV)}$~\cite{Baur:2002qd,Baur:2003gp}. 
At a linear collider the Higgs self coupling, $g_{hhh}$, can be accessed by measuring the 
cross section of the $h^0 h^0 Z^0$ and $h^0 h^0 \nu \bar \nu$ processes from 0.5 to 
3~TeV~\cite{Battaglia:2001nn}. 
Isolating the $h^0 h^0 Z^0$ signal ($\sigma$=0.18~fb at $\sqrt{s}$ = 0.5~TeV) from $t \bar t$ 
(530~fb), $Z^0 Z^0 Z^0$ (1.1~fb), $t \bar b \bar t b$ (0.7~fb) and $t\ \bar t Z^0$ (1.1~fb)  
will be a genuine experimental tour-de-force. Again, $b$ tagging is essential for background 
suppression, provided high efficiency is attained, given the tiny
signal cross section. Preliminary studies performed at 0.5~TeV for the $h^0 h^0 Z^0$ channel have 
shown that once a realistic simulation of signal and backgrounds is performed the results
are less optimistic, compared to those obtained with parametric simulations. A multi-TeV 
collider is in principle advantageous for this measurement, since the cross section for the 
$h^0 h^0 \nu \bar \nu$ process at 3~TeV is larger by a factor of seven and can be further 
increased by operating with polarised beams. Still the $W^+ W^- \nu \bar \nu$ (125~fb), 
$Z^0 Z^0 \nu \bar \nu$ (54~fb), $W^+ W^- Z^0$ (32~fb) and $Z^0 Z^0 Z^0$ (0.34~fb) backgrounds are 
significant. Further, most of the sensitivity to $g_{hhh}$ is in the forward region, 
which poses challenges in terms of reconstruction.

Moving away from the Higgs sector, $b$ quarks remain a leading signature for several scenarios 
of new physics. A class of SUSY models with large hierarchy between the scalars and the gauginos, 
the so-called split Supersymmetry, have recently been studied in detail~\cite{ArkaniHamed:2004yi}. 
In these models quarks and sleptons are heavy enough that charginos and neutralinos 
decay exclusively into $W^{\pm}$, $Z^0$ and $h^0$ bosons and lighter $\chi$ states. In particular, 
decays of the kind $\chi^0_{2,3,4} \to h^0 \chi^0_{1,2}$ and  $\chi^{\pm}_{2} \to h^0 \chi^{\pm}_1$, 
followed by $h^0 \to b \bar b$ may be dominant~\cite{Bernal:2007uv}. These features are also common 
to mSUGRA models with large values of $m_0$, $m_{1/2}$ and $\tan \beta$. All these give remarkable 
events with four $b$-jets and large missing energy, which would be the gaugino sector counterpart 
of the events from heavy Higgs decays discussed above. 

The study of precision electro-weak observables in two-fermion production opens an window 
on phenomena at mass scales well above the collider collision energy. The $e^+e^- \to b \bar b$ 
and $e^+e^- \to t \bar t$ processes are important in this respect because they allow us to 
select samples of down- and up-type quarks selected with good purity. In addition, models of 
warped extra dimensions with bulk SM fields have excitations strongly coupled to quarks of the third 
generation. These excitation may be heavier that the $\sqrt{s}$ energy of the 
collider~\cite{Contino:2006nn}. In this case, deviations to the cross section and forward-backward 
asymmetries in the $b \bar b$ and $t \bar t$ two-fermion processes, should be detectable at a 
multi-TeV linear collider up to masses of $\cal{O} \mathrm{(10~TeV)}$ 

In summary, the $b$- and $c$-tagging efficiency has to be large for most of the anticipated studies, 
which are characterised by large jet multiplicity and low signal-to-background ratios. The systematics 
coming from input parameters in the theory predictions and the efficiency of the jet flavour 
tagging also need to be minimised. Heavy quark masses play a major role in the interpretation 
of the Higgs sector data. Heavy hadron production and decay properties, in particular charged decay 
multiplicities ands fragmentation functions, are crucial to jet flavour tagging, which is based on 
the topology and kinematics of charged particle tracks. For this part, the linear collider program 
will largely rely on results obtained at lower energy heavy flavour experiments.

\section{VERTEX TRACKERS FOR HEAVY FLAVOUR TAGGING}

The crucial role of heavy flavour tagging at a linear collider has motivated the 
significant attention that the design of a vertex tracker and the choice of sensor 
technologies has received in the last decade. Vertex trackers at a future $e^+e^-$ 
linear collider will face new and different challenges compared to those at LEP and LHC. 
The distinctive feature of linear collider physics is its anticipated accuracy for a large 
variety of measurements (spectroscopy, searches, rare decays, electro-weak observables, ...) 
to be performed over a broad energy range. 
The target performance for the track extrapolation resolution is $\sigma_{IP} = 
5 {\mathrm{\mu m}} \oplus \frac{10 {\mathrm{\mu m~GeV^{-1}}}}{p_t}$ for operation
below 1~TeV and  $5 {\mathrm{\mu m}} \oplus \frac{20 {\mathrm{\mu m~GeV^{-1}}}}{p_t}$ 
for a multi-TeV collider, where the larger particle boost and energy is expected to 
compensate at least in part the larger distance of the detector from the beam, dictated 
by beam-induced backgrounds. This performance makes possible the detailed reconstruction 
of the decay topology in hadronic jets containing a $b$ or a $c$ quark. This reconstruction 
allows us to identify not only $b$ jets with high efficiency, $\epsilon_b$ = 0.80 with a 
misidentification probability of 0.02 and 0.25 for $u$, $d$, $s$ and $c$ quarks, respectively, 
but also $c$ jets with high purity, even when the main background consists of $b$ jets, 
$\epsilon_c$ = 0.30 with 0.07 misidentification probability for $b$~\cite{Bailey:2009ui}. 
This is the case of the $h^0 \to c \bar c$ decay for $M_h$ = 120~GeV, where the dominant 
$h^0 \to b \bar b$ decays constitute the main background, giving a signal-to-background 
ratio of 0.05. The asymptotic track extrapolation resolution can be obtained with a single 
point resolution of $\simeq$~3~$\mu$m and the multiple scattering term implies a single 
layer thickness of $\simeq$~0.1~\%~$X_0$. The single point resolution corresponds to a pixel 
size of 10-30~$\mu$m with binary or analog readout. The layer material budget requires 
$\simeq$ 50~$\mu$m-thick sensors and puts significant constraints on the chip power dissipation, 
which should be kept compatible with passive cooling. Data can be read-out either continuously 
during the train of colliding bunches to keep the detector occupancy low, corresponding to a 
readout
time of 25-50~$\mu$s at the ILC, or stored locally with a time stamp and read-out in between 
trains, depending on the pixel granularity and the read-out architecture chosen. In the case 
of CLIC, where the bunch spacing is only 0.5~ns, time stamping to $\simeq$10-20~ns is likely 
required to keep the occupancy low and to identify tracks from 
$\gamma \gamma \to \mathrm{hadrons}$ 
produced outside of the time bucket of the main $e^+e^-$ collision of interest.

\begin{figure}[htb]
\includegraphics[width=17pc]{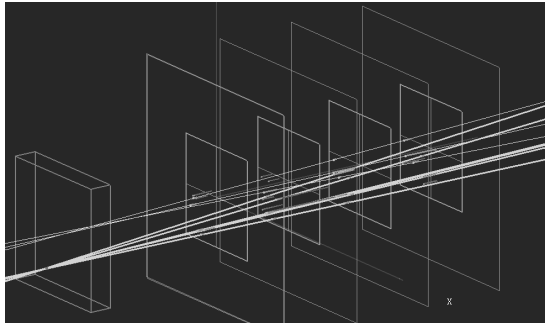}
\vspace*{-0.5cm}
\caption{Display of a six-prong vertex from the interaction of an 120~GeV proton 
in a Cu target reconstructed with a beam hodoscope made of thin CMOS pixel sensors
in the T966 beam test experiment at FNAL. The average resolution of the vertex position 
along the beam axis has been measured to be (260$\pm$10)~$\mu$m, which corresponds to 
the vertexing resolution expected for a detector at CLIC (from~\cite{Battaglia:2008nj}).}
\label{fig:t966}
\end{figure}

Radiation conditions are significantly lower compared to those faced by the LHC detectors 
and novel sensor technologies can be  exploited. 
The main path of R\&D to match the linear collider requirements is towards detectors which 
have substantially lower material budget and higher space, or space-time granularity,  
compared to those developed for the LHC. This can be achieved with monolithic 
technologies, where the sensitive volume and at least part of the signal processing 
electronics are implemented in the same Si wafer. CMOS active pixel sensors have 
demonstrated several appealing properties and have been adopted as baseline for detailed 
designs~\cite{DeMasi:2010}.
Beam hodoscopes made of pixel sensors of various technologies, developed for linear collider 
application  have been successfully operated and demonstrated tracking performances meeting 
the LC requirements under realistic conditions~\cite{Battaglia:2008nj,Velthuis:2008zza,eudet,JBaudot}  
(see Figure~\ref{fig:t966}).
Some of these technologies have now reached a degree of maturity and reliability that 
makes them well suited for application in near-term projects. Vertex detector based on pixel 
sensors, originally developed for the ILC are currently under construction for the HFT upgrade 
of STAR at RHIC~\cite{hft} and for Belle-2 at KEKB~\cite{andricek}, based on the CMOS and 
DEPFET technologies, respectively. 
However, these still have limitations in readout speed and in amount of data processing 
performed directly in pixel. New technologies, potentially more performant and which 
may accommodate more advanced functionalities have recently emerged. Silicon-On-Insulator (SOI) 
with a high resistivity handle wafer brings together electronics in low feature size CMOS 
processese with a fully depleted sensitive substrate~\cite{Arai,Battaglia:2008rt}.   
Multi-tier vertical integration techniques make possible to integrate heterogeneous 
technologies by stacking several thin layers with small pitch interconnect. This gives 
maximum freedom of choice for the use of optimal technologies for the various functions
of the detector~\cite{yarema}. Both technologies are moving through the stages of a generic 
R\&D exploring feasibility and issues of their use for particle detection and design 
optimisation. 

Sensor R\&D motivated by the linear collider program has already found important 
applications in other project in HEP as well as fields of science outside of accelerator 
particle physics. Thin CMOS sensors with small, radiation tolerant pixels are successfully used 
in imaging in transmission electron microscopy, where their thin sensitive volume and high 
pixellisation is very advantageous and provides point spread function performances superior to 
photographic films at a frame rate of $\simeq$~100~f/s~\cite{tem}. DEPFET sensors have excellent 
energy resolution and are being developed for soft X-ray detection in planetary 
imaging~\cite{mixs} and XFEL~\cite{xfel} experiments.


\begin{thebibliography}{99}

\bibitem{Brau:2007zza}
  J.~Brau {\it et al.}, SLAC-R-857 (2007).

\bibitem{Assmann:2000hg}
  R.~W.~Assmann {\it et al.}, CERN 2000-008.

\bibitem{Geddes:2004tb}
  C.~G.~R.~Geddes {\it et al.},
  Nature {\bf 431} (2004) 538.

\bibitem{Thomas:2007jg}
  A.~G.~R.~Thomas {\it et al.},
  Phys.\ Rev.\ Lett.\  {\bf 98} (2007) 095004
  [arXiv:physics/0701186].

\bibitem{Battaglia:2000jb}
  M.~Battaglia and K.~Desch,
  AIP Conf.\ Proc.\  {\bf 578} (2001) 163
  [arXiv:hep-ph/0101165].

\bibitem{Bailey:2009ui}
  D.~Bailey {\it et al.}  [LCFI Collaboration],
  Nucl.\ Instrum.\ Meth.\  A {\bf 610} (2009) 573
  [arXiv:0908.3019 [physics.ins-det]].

\bibitem{Kuhl:2007zza}
  T.~Kuhl and K.~Desch, Note LC-PHSM-2007-001.

\bibitem{ildloi}
  H.~Stoeck {\it et al.}, The ILD Letter of Intent, 2009.

\bibitem{Battaglia:2002gq}
  M.~Battaglia and A.~De Roeck,
  arXiv:hep-ph/0211207.

\bibitem{Barklow:2003hz}
  T.~L.~Barklow,
  arXiv:hep-ph/0312268.

\bibitem{Droll:2006se}
  A.~Droll and H.~E.~Logan,
  Phys.\ Rev.\  D {\bf 76} (2007) 015001
  [arXiv:hep-ph/0612317].

\bibitem{Battaglia:2010in}
  M.~Battaglia and P.~Ferrari, CERN-LCD-2010-006, 
  arXiv:1006.5659 [hep-ex].

\bibitem{Christova:2002ke}
  E.~Christova {\it et al.},
  Nucl.\ Phys.\  B {\bf 639} (2002) 263
  [Erratum-ibid.\  B {\bf 647} (2002) 359]
  [arXiv:hep-ph/0205227].

\bibitem{Baur:2002qd}
  U.~Baur, T.~Plehn and D.~L.~Rainwater,
  Phys.\ Rev.\  D {\bf 67} (2003) 033003
  [arXiv:hep-ph/0211224].

\bibitem{Baur:2003gp}
  U.~Baur, T.~Plehn and D.~L.~Rainwater,
  Phys.\ Rev.\  D {\bf 69} (2004) 053004
  [arXiv:hep-ph/0310056].

\bibitem{Battaglia:2001nn}
  M.~Battaglia, E.~Boos and W.~M.~Yao,
  in {\it Proc. of the APS/DPF/DPB Summer Study on the Future of Particle Physics (Snowmass 2001) } 
  ed. N.~Graf, E3016,  [arXiv:hep-ph/0111276].

\bibitem{ArkaniHamed:2004yi}
  N.~Arkani-Hamed {\it et al.},
  Nucl.\ Phys.\  B {\bf 709} (2005) 3
  [arXiv:hep-ph/0409232].

\bibitem{Bernal:2007uv}
  N.~Bernal, A.~Djouadi and P.~Slavich,
  JHEP {\bf 0707} (2007) 016
  [arXiv:0705.1496 [hep-ph]].

\bibitem{Contino:2006nn}
  R.~Contino {\it et al.},
  JHEP {\bf 0705} (2007) 074
  [arXiv:hep-ph/0612180].

\bibitem{DeMasi:2010}
  R.~De~Masi {\it et al.},
  to appear on Nucl.\ Instrum.\ Meth.\  A (2010), doi:10.1016/j.nima.2010.06.339.

\bibitem{Battaglia:2008nj}
  M.~Battaglia {\it et al.},
  Nucl.\ Instrum.\ Meth.\  A {\bf 593} (2008) 292
  [arXiv:0805.1504 [physics.ins-det]].

\bibitem{Velthuis:2008zza}
  J.~J.~Velthuis {\it et al.},
  IEEE Trans.\ Nucl.\ Sci.\  {\bf 55} (2008) 662.

\bibitem{eudet}
  P.~Roloff,
  Nucl.\ Instrum.\ Meth.\  A {\bf 604} (2009) 265.

\bibitem{JBaudot}
  J.~Baudot {\it et al.},
  IEEE Trans.\ Nucl.\ Sci.\  {\bf 56} (2009) 1677.

\bibitem{hft}
  L.~Greiner {\it et al.},
  Nucl.\ Instrum.\ Meth.\  A {\bf 589} (2007) 675.

\bibitem{andricek}
  L.~Andricek {\it et al.},
  to appear on Nucl.\ Instrum.\ Meth.\  A (2010), doi:10.1016/j.nima.2010.02.191.

\bibitem{Arai}
  Y.~Arai {\it et al.},
  to appear on Nucl.\ Instrum.\ Meth.\  A (2010), doi:10.1016/j.nima.2010.02.190.

\bibitem{Battaglia:2008rt}
  M.~Battaglia {\it et al.},
  Nucl.\ Instrum.\ Meth.\  A {\bf 604} (2009) 380
  [arXiv:0811.4540 [physics.ins-det]].

\bibitem{yarema}
  R.~Yarema {\it et al.},
  Nucl.\ Instrum.\ Meth.\  A {\bf 617} (2010) 375.

\bibitem{tem}
  M.~Battaglia {\it et al.},
  to appear on Nucl.\ Instrum.\ Meth.\  A (2010), doi:10.1016/j.nima.2010.07.066.

\bibitem{mixs}
  J.~Treis {\it et al.},
  to appear on Nucl.\ Instrum.\ Meth.\  A (2010), doi:10.1016/j.nima.2010.03.173.

\bibitem{xfel}
  M.~Porro {\it et al.},
  to appear on Nucl.\ Instrum.\ Meth.\  A (2010), doi:10.1016/j.nima.2010.02.254.

\end{thebibliography}
\end{document}